\documentclass[11pt]{article}
\usepackage[a4paper,margin=1in]{geometry}
\usepackage{amsmath}
\usepackage{graphicx}
\usepackage{booktabs}
\usepackage[font=small,labelfont=bf]{caption}
\usepackage[hidelinks]{hyperref}
\graphicspath{{figs/}}

\title{Adaptive Query Routing: A Tier-Based Framework for Hybrid Retrieval Across Financial, Legal, and Medical Documents}
\author{Afshan Hashmi\\
Tuwaiq Research and Development Center (TRDC),\\ Tuwaiq Academy, Riyadh, Saudi Arabia\\
\texttt{afshanhashmi786@gmail.com} $|$ \texttt{a.hashmi@tuwaiq.edu.sa}}
\date{October 2026}

\begin{document}
\maketitle

\begin{abstract}
Retrieval-Augmented Generation (RAG) has become the standard paradigm for grounding Large Language Model outputs in external knowledge. Lumer et al.~\cite{lumer} presented the first systematic evaluation comparing vector-based agentic RAG against hierarchical node-based reasoning systems for financial document QA across 1,200 SEC filings, finding vector-based systems achieved a 68\% win rate. Concurrently, the PageIndex framework~\cite{pageindex} reported 98.7\% accuracy on FinanceBench through purely reasoning-based retrieval. This paper extends their work by: (i) implementing and evaluating three retrieval configurations: Vector RAG, Tree Reasoning, and the proposed Adaptive Hybrid Retrieval (AHR), on a small controlled corpus of financial, legal, and medical documents; (ii) introducing a four-tier query complexity benchmark; and (iii) employing LLM-as-judge evaluation with GPT-4o-mini. Experiments reveal that Tree Reasoning achieves the highest overall score (0.900), but no single paradigm dominates across all tiers: AHR scores highest on cross-reference queries (0.850), all three methods tie on multi-section queries (0.929), and Vector RAG scores highest on the single multi-document query (0.900). Vector RAG has the lowest mean latency (1.61\,s against 3.40\,s for Tree Reasoning). On the first 50 publicly released FinanceBench questions, using the evidence snippets supplied with the benchmark as the retrieval corpus, Tree Reasoning scores 0.938, Hybrid AHR 0.901, and Vector RAG 0.821. Because the configurations also differ in document access, generation prompts and output length limits, these differences cannot be attributed to retrieval architecture alone. In these preliminary results the adaptive router does not outperform fixed tree reasoning overall (0.845 vs.\ 0.900); they motivate a larger evaluation of when query-based routing helps, rather than establishing its benefit.
\end{abstract}


\section{Introduction}
Large Language Models (LLMs) have transformed document understanding, yet remain constrained by context window limitations~\cite{liu,gao}. Retrieval-Augmented Generation (RAG) emerged as the dominant solution, with vector databases becoming the default infrastructure~\cite{lewis,dpr}. However, semantic similarity does not reliably capture contextual relevance~\cite{pageindex,sarkar}, particularly for structured professional documents.

Lumer et al.~\cite{lumer} published the first systematic comparison of vector-based and non-vector RAG for financial QA across 1,200 SEC filings with 150 expert-curated questions, finding vector-based agentic RAG achieved a 68\% win rate over hierarchical node-based systems. In the same study, cross-encoder reranking~\cite{nogueira,sun} raised MRR@5 from 0.160 to 0.750~\cite{lumer}. Meanwhile, PageIndex~\cite{pageindex} reported 98.7\% accuracy on FinanceBench through tree-based reasoning without any vector database.

These seemingly contradictory findings highlight open questions: (1) whether the comparison carries over to other domains, (2) when each paradigm excels, and (3) whether an adaptive system can combine their complementary strengths. This paper takes a first, small-scale step on all three.

\paragraph{Contributions.} The contributions of this paper are as follows:
\begin{enumerate}
\item A cross-domain implementation comparing Vector RAG, Tree Reasoning RAG, and a proposed Adaptive Hybrid Retrieval (AHR) framework on financial, legal, and medical documents.
\item A four-tier query complexity benchmark testing simple lookups through cross-reference following, with GPT-4o-mini LLM-as-judge evaluation of answer quality and measurement of per-query latency.
\item Experimental evidence that no single paradigm dominates all query types, with AHR scoring highest on cross-reference queries.
\item A supplementary evaluation on the first 50 of FinanceBench's~\cite{financebench} 150 publicly released questions on real SEC 10-K and 10-Q filings, using the benchmark's supplied evidence snippets, in which, as on the controlled corpus, Tree Reasoning obtains the highest mean judge score.
\end{enumerate}

\section{Related Work}
\subsection{Evolution of RAG Architectures}
The RAG paradigm has evolved through three generations~\cite{gao,ranjan}. Naive RAG (2020--2022) established the retrieve-then-generate pipeline~\cite{lewis}. Advanced RAG introduced query rewriting~\cite{ma}, cross-encoder reranking~\cite{nogueira,sun}, and corrective RAG~\cite{yan}. Modular RAG decomposed the pipeline into interchangeable components~\cite{hu}. Adaptive-RAG~\cite{adaptiverag} selects among no retrieval, single-step, and multi-step retrieval according to predicted question complexity; the router studied here instead selects between retrieval \emph{types}. Recent surveys provide comprehensive coverage~\cite{wu,zhao,sharma,yu}.

\subsection{Vector-Based Retrieval}
Dense passage retrieval (DPR)~\cite{dpr} established embedding-based search using dual-encoder architectures. ColBERT~\cite{colbert} introduced late-interaction retrieval, while hybrid approaches combined BM25 with neural retrievers~\cite{bm25}. For financial documents specifically, Setty et al.~\cite{setty} introduced metadata-enhanced retrieval, and RAG-Fusion~\cite{ragfusion} improved recall through reciprocal rank fusion.

\subsection{Non-Vector and Structured Retrieval}
PageIndex~\cite{pageindex} introduced hierarchical tree indexing with LLM-powered tree search, inspired by AlphaGo~\cite{alphago}. GraphRAG~\cite{graphrag} represented documents as knowledge graphs. HiRAG~\cite{hirag} introduced hierarchical knowledge retrieval at EMNLP 2025. Graph-of-Thought~\cite{got} structured LLM reasoning as graphs, and KRAGEN~\cite{kragen} used graph-of-thoughts prompting for subproblem decomposition.

\subsection{Financial Domain RAG}
TAT-QA~\cite{tatqa} established hybrid tabular-textual benchmarks. FinanceBench~\cite{financebench} introduced 10,231 expert-annotated questions on real SEC filings, 150 of which are publicly released; the original evaluation found that GPT-4 Turbo with standard RAG answered only 19\% of questions correctly, demonstrating the benchmark's difficulty. Setty et al.~\cite{setty} subsequently improved to 25.6\% through metadata-enhanced retrieval, and Jimeno-Yepes et al.~\cite{jimeno} reached 32.6\% with improved chunking strategies. FinSage~\cite{finsage} represents the current state-of-the-art among vector-based systems on FinanceBench, achieving 49.66\% LLM accuracy through multi-path retrieval with domain-specialised reranking. RLFKV~\cite{rlfkv} addressed hallucination through fine-grained knowledge verification. RankRAG~\cite{rankrag} unified context ranking with generation at NeurIPS 2024.

\section{Methodology}
\subsection{Multi-Domain Document Dataset}
A controlled evaluation corpus of four documents was constructed spanning three domains: two financial documents (an SEC-style 10-K annual report with five sections and three subsections, including a risk-quantification appendix, and a 10-Q quarterly report with four sections), one legal document (a master services agreement with eight sections and two subsections, including definitions and exhibits), and one medical document (a Phase III clinical study report with five sections and three subsections covering efficacy, safety, and subgroup analyses). Sections contain explicit cross-references (e.g., ``See Appendix A for detailed risk quantification''). The documents are short: Vector RAG indexes 30 chunks in total.

\subsection{Four-Tier Query Complexity Benchmark}
Following the query complexity analysis implicit in Lumer et al.~\cite{lumer}, a four-tier classification is formalised (Table~\ref{tab:tiers}). The benchmark comprises 22 queries (10 financial, 6 legal, 6 medical): 8 in Tier~1, 7 in Tier~2, 6 in Tier~3, and 1 in Tier~4. Each query has a ground-truth answer and a list of the sections that contain its evidence.

\begin{table}[ht]
\centering
\caption{Query complexity tier classification.}
\label{tab:tiers}
\begin{tabular}{lllll}
\toprule
Tier & Type & Proposed routing strategy & Example & $n$ \\
\midrule
T1 & Single-fact lookup & Vector similarity & What was Q3 revenue? & 8 \\
T2 & Multi-section reasoning & Tree navigation & Revenue vs.\ guidance? & 7 \\
T3 & Cross-reference & Hierarchical traversal & Appendix details? & 6 \\
T4 & Multi-document synthesis & Hybrid fusion & Compare across filings & 1 \\
\bottomrule
\end{tabular}
\end{table}

\subsection{System Implementations}
\paragraph{Vector RAG.} Each section and subsection is split into chunks of 150 words with a 30-word overlap (chunks of 10 words or fewer are discarded), embedded using all-MiniLM-L6-v2~\cite{sbert}, and indexed with FAISS~\cite{faiss}. The top-5 chunks across the whole corpus are retrieved via cosine similarity. GPT-4o-mini generates answers from the retrieved context, prompted for a concise answer with a 300-token output limit.

\paragraph{Tree Reasoning RAG.} Each document is parsed into a tree of sections and subsections, and GPT-4o-mini writes a one- or two-sentence summary for every node. At query time, GPT-4o-mini reads the tree of the query's target document and returns the section identifiers it judges relevant; the content of those sections is retrieved. Explicit references in the retrieved text (``See Appendix~A'', ``as specified in Exhibit~B'') are detected with regular expressions and the referenced sections are added if not already selected. Unlike Vector RAG, which searches all documents, Tree Reasoning is given the query's target document (for the multi-document query, the annual report). Its answers are generated with a prompt requesting a detailed answer and a 400-token output limit.

\paragraph{Adaptive Hybrid RAG (Proposed).} A GPT-4o-mini query classifier assigns each query to a complexity tier. Tier 1 queries route to vector search; Tier 2--3 queries route to tree reasoning on the target document; Tier 4 queries invoke vector search (top-8) and tree reasoning over every document, and GPT-4o-mini fuses the partial answers.

\subsection{Evaluation Protocol}
Following Lumer et al.~\cite{lumer}, LLM-as-judge evaluation is employed~\cite{zheng} using GPT-4o-mini. Given the question, the ground-truth answer, and the system answer, the judge scores accuracy, completeness, and relevance, and gives an overall score, each on a 0--10 scale rescaled to 0--1; the overall score is reported as answer quality. Latency covers per-query retrieval or tree navigation and answer generation, including the routing call for the Hybrid AHR; it excludes offline indexing, tree construction and judging. Each configuration was run once. Section-level retrieval metrics are not reported: the original scoring code matched section numbers without the identity of the document they belong to, so a same-numbered section from another document could be counted as correct, and the saved outputs do not retain the document-qualified identifiers needed to recompute them. The judge ratings and latencies were computed independently of that scoring step.

\subsection{FinanceBench Evaluation}
As a supplementary evaluation on real documents, all three configurations were evaluated on FinanceBench~\cite{financebench}, a publicly available benchmark of 150 expert-annotated question-answer pairs drawn from real SEC 10-K and 10-Q filings. The first 50 questions of the released file were used; they cover 10 of its 32 companies. FinanceBench supplies, for each question, the evidence snippets that annotators extracted from the filings. The retrieval corpus is built from these snippets for all 84 filings in the benchmark, grouped by filing and organised into sections labelled with their cited page numbers; it therefore contains neither complete pages nor complete filings. Vector RAG indexes 200-word chunks of this text (270 chunks), retrieves the 15 nearest chunks, and keeps up to five from the question's filing; Tree Reasoning presents GPT-4o-mini with the snippet sections of the question's filing, each with its page label and first 120 characters, and reads the sections it selects. The router classifies each question as SIMPLE (sent to vector search) or STRUCTURAL (sent to tree reasoning). The judge uses a separate prompt that scores accuracy, completeness, and relevance on a 0--1 scale and reports their mean as the overall score. Because the corpus, router, and judge differ from the controlled setting, FinanceBench scores are compared with controlled-corpus scores only in direction, not in magnitude.

\section{Experimental Results}
\subsection{Overall Performance}
Table~\ref{tab:overall} presents aggregate performance across all 22 queries and three domains.

\begin{table}[ht]
\centering
\caption{LLM-as-judge ratings and latency on the controlled corpus (22 queries, 3 domains).}
\label{tab:overall}
\small\setlength{\tabcolsep}{4pt}
\begin{tabular}{lccccc}
\toprule
Method & Quality & Accuracy & Complete. & Relevance & Latency \\
\midrule
Tree Reasoning & \textbf{0.900} & \textbf{0.936} & \textbf{0.845} & \textbf{0.968} & 3.40\,s \\
Hybrid AHR & 0.845 & 0.886 & 0.755 & 0.936 & 3.10\,s \\
Vector RAG & 0.845 & 0.877 & 0.759 & 0.950 & \textbf{1.61\,s} \\
\bottomrule
\end{tabular}
\end{table}

Tree Reasoning achieves the highest quality (0.900), and Vector RAG and Hybrid AHR tie at 0.845; Tree Reasoning also has the highest accuracy, completeness and relevance ratings. Vector RAG maintains the lowest latency at 1.61\,s, about half that of Tree Reasoning (3.40\,s). The quality gap should not be read as an effect of retrieval architecture alone: Tree Reasoning is given the target document and generates longer, more detailed answers (Section~3.3), which may also raise its completeness ratings.

\subsection{Performance by Query Complexity Tier}
\begin{figure}[ht]
\centering
\includegraphics[width=0.85\textwidth]{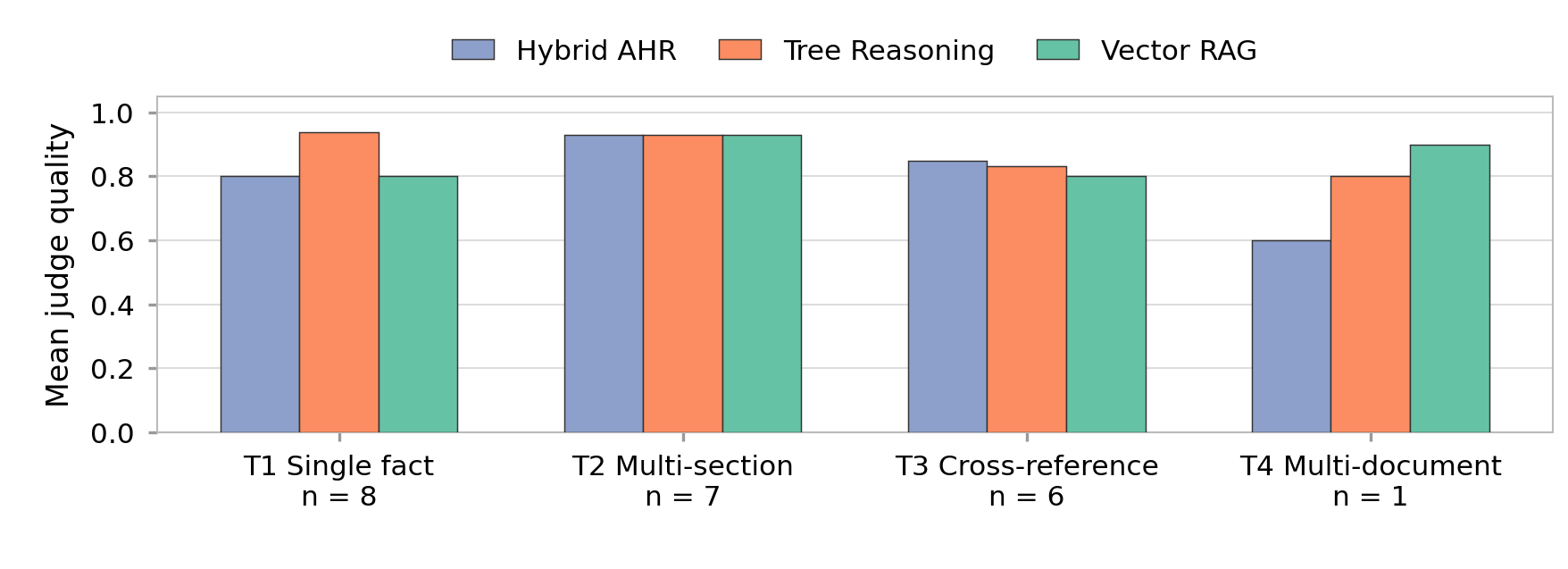}
\caption{Mean LLM-as-judge quality by query complexity tier on the controlled corpus ($n$ = 8, 7, 6 and 1 queries).}
\label{fig:tier}
\end{figure}

Table~\ref{tab:tier} and Figure~\ref{fig:tier} show that no single method dominates across all complexity tiers. Tree Reasoning leads on simple queries (T1: 0.938). All three methods tie on multi-section queries (T2: 0.929). Hybrid AHR achieves the highest score on cross-reference queries (T3: 0.850). Vector RAG scores highest on Tier~4 (0.900), which contains a single query. Hybrid AHR's Tier~1 score equals Vector RAG's, consistent with the router sending these queries to vector search.

\begin{table}[ht]
\centering
\caption{LLM-as-judge quality by complexity tier (controlled corpus).}
\label{tab:tier}
\begin{tabular}{lcccc}
\toprule
Tier & $n$ & Vector RAG & Tree Reasoning & Hybrid AHR \\
\midrule
T1 Single-fact lookup & 8 & 0.800 & \textbf{0.938} & 0.800 \\
T2 Multi-section reasoning & 7 & 0.929 & 0.929 & 0.929 \\
T3 Cross-reference & 6 & 0.800 & 0.833 & \textbf{0.850} \\
T4 Multi-document synthesis & 1 & \textbf{0.900} & 0.800 & 0.600 \\
\bottomrule
\end{tabular}
\end{table}

\subsection{Cross-Domain Analysis}
Table~\ref{tab:domain} presents domain-specific answer quality.

\begin{table}[ht]
\centering
\caption{LLM-as-judge quality score by domain.}
\label{tab:domain}
\begin{tabular}{lcccc}
\toprule
Domain & Queries & Vector RAG & Tree Reasoning & Hybrid AHR \\
\midrule
Financial & 10 & \textbf{0.910} & 0.890 & 0.880 \\
Legal & 6 & 0.800 & \textbf{0.883} & 0.850 \\
Medical & 6 & 0.783 & \textbf{0.933} & 0.783 \\
\bottomrule
\end{tabular}
\end{table}

Vector RAG has the highest mean score in the financial domain (0.910), and Tree Reasoning in the legal (0.883) and medical (0.933) domains. The ratings alone do not show why; they do not separate the contribution of structural navigation from the other differences between configurations. With one legal and one medical document of six queries each, these domain differences are illustrative rather than general.

\subsection{Latency vs.\ Quality Trade-off}
\begin{figure}[ht]
\centering
\includegraphics[width=0.7\textwidth]{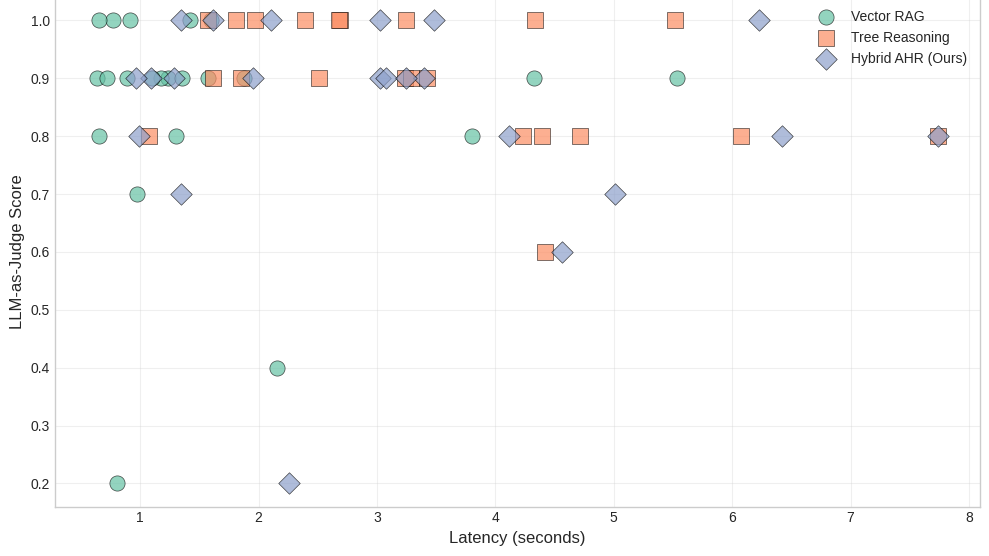}
\caption{Latency vs.\ answer quality for all 66 query--method pairs on the controlled corpus.}
\label{fig:scatter}
\end{figure}

Figure~\ref{fig:scatter} shows the latency and quality of all 66 query--method pairs on the controlled corpus. Vector RAG answers most queries in under 2\,s but its quality varies widely (0.2--1.0), while Tree Reasoning scores between 0.6 and 1.0 at roughly 1--8\,s. Hybrid AHR achieves an intermediate profile, reducing average latency from 3.40\,s to 3.10\,s, but its lowest score (0.2) is as low as Vector RAG's.

\subsection{Cross-Reference Resolution}
Across the six Tier~3 cross-reference queries the three systems are close in answer quality, with Hybrid AHR at 0.850, Tree Reasoning at 0.833 and Vector RAG at 0.800 (Table~\ref{tab:tier}). These are answer ratings, not measurements of how much referenced evidence was retrieved, and with six queries the differences are small. In the recorded runs, the regular-expression reference-following step never added a section beyond those the tree navigator had already selected; whether it helps would require an ablation under matched conditions.

\subsection{Results by Domain and Tier}
\begin{figure}[ht]
\centering
\includegraphics[width=\textwidth]{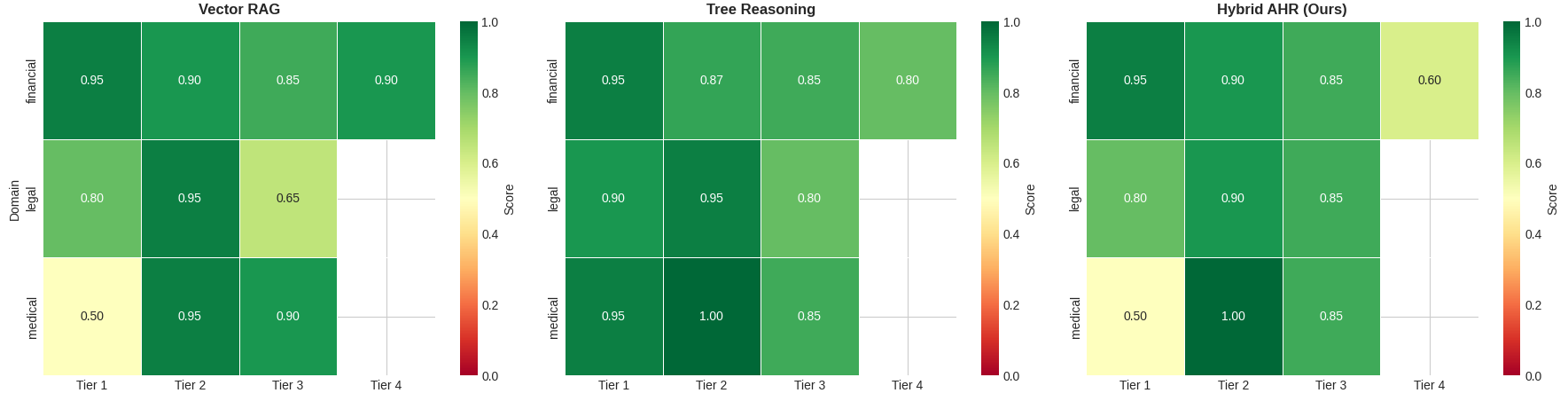}
\caption{Answer quality by domain and tier for each retrieval method. Empty cells have no queries.}
\label{fig:heatmap}
\end{figure}

Figure~\ref{fig:heatmap} shows how the observed scores vary across domain and tier; no formal interaction analysis was performed. For medical Tier~1 queries, Vector RAG and Hybrid AHR average 0.50 while Tree Reasoning averages 0.95, and for medical Tier~2 queries Tree Reasoning and Hybrid AHR both average 1.00. With so few queries per cell, these figures describe the runs rather than explain them.

\subsection{FinanceBench Results}
Table~\ref{tab:fb} presents results on FinanceBench ($n=50$, real SEC 10-K and 10-Q filings). As on the controlled corpus, Tree Reasoning has the highest mean judge score; unlike the controlled corpus, Hybrid AHR is here ahead of Vector RAG rather than tied with it.

\begin{table}[ht]
\centering
\caption{Performance on FinanceBench ($n=50$; separate judge prompt, see Section~3.5).}
\label{tab:fb}
\begin{tabular}{lccccc}
\toprule
Method & Overall & Accuracy & Completeness & Relevance & Latency \\
\midrule
Tree Reasoning & \textbf{0.938} & \textbf{0.936} & \textbf{0.914} & \textbf{0.964} & 4.15\,s \\
Hybrid AHR (Proposed) & 0.901 & 0.884 & 0.876 & 0.942 & 4.36\,s \\
Vector RAG & 0.821 & 0.798 & 0.768 & 0.898 & \textbf{2.40\,s} \\
\bottomrule
\end{tabular}
\end{table}

Tree Reasoning scores above Vector RAG in both evaluations, but the size of the gap cannot be compared between them, since the judging protocol and retrieval setting differ. Part of Vector RAG's deficit likely reflects its retrieval setting: it searches all filings and then keeps only chunks from the question's filing, so for some questions few or no chunks remain, whereas Tree Reasoning is given the filing directly. The comparison therefore reflects document access as well as retrieval method. The Hybrid AHR routed 74\% of FinanceBench queries to tree reasoning and 26\% to vector search, compared to 55\%/45\% on the controlled corpus.

\begin{figure}[ht]
\centering
\includegraphics[width=0.85\textwidth]{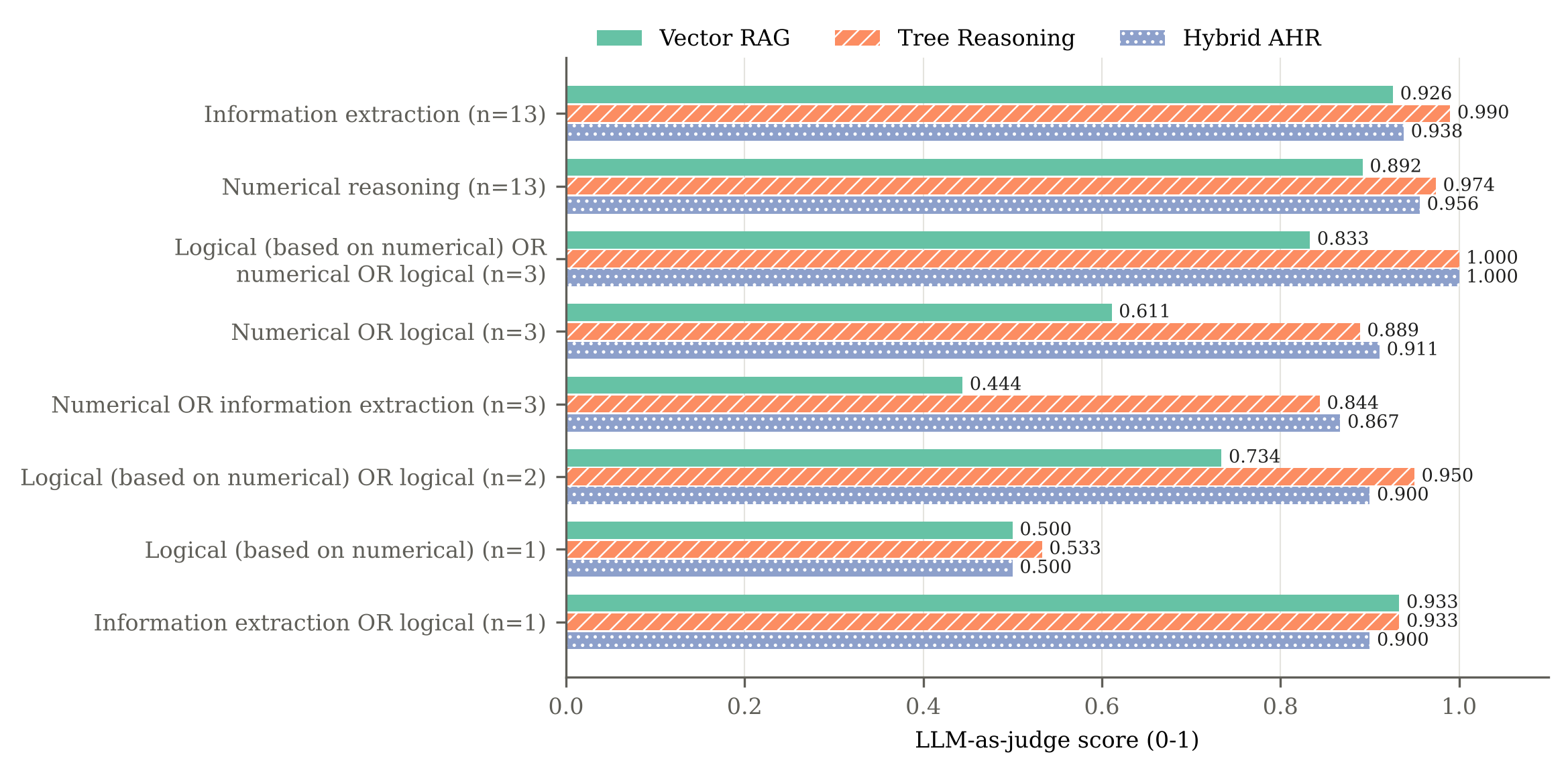}
\caption{FinanceBench quality by reasoning label ($n$ = questions per label).}
\label{fig:qtype}
\end{figure}

Table~\ref{tab:qtype} and Figure~\ref{fig:qtype} break down performance by FinanceBench's reasoning labels; 11 of the 50 questions carry no label and are omitted. Tree Reasoning has its highest mean on information extraction questions (0.990). On the one question labelled as logical reasoning based on numerical reasoning, the three systems score between 0.500 and 0.533. Vector RAG falls furthest behind on questions with mixed labels.

\begin{table}[ht]
\centering
\caption{FinanceBench performance by FinanceBench reasoning label (labels as released).}
\label{tab:qtype}
\small
\begin{tabular}{lcccc}
\toprule
Reasoning label & $n$ & Vector RAG & Tree Reasoning & Hybrid AHR \\
\midrule
Information extraction & 13 & 0.926 & \textbf{0.990} & 0.938 \\
Numerical reasoning & 13 & 0.892 & \textbf{0.974} & 0.956 \\
Logical (num.) OR Numerical OR Logical & 3 & 0.833 & \textbf{1.000} & \textbf{1.000} \\
Numerical OR Logical & 3 & 0.611 & 0.889 & \textbf{0.911} \\
Numerical OR Information extraction & 3 & 0.444 & 0.844 & \textbf{0.867} \\
Logical (num.) OR Logical & 2 & 0.734 & \textbf{0.950} & 0.900 \\
Logical (based on numerical reasoning) & 1 & 0.500 & \textbf{0.533} & 0.500 \\
Information extraction OR Logical & 1 & \textbf{0.933} & \textbf{0.933} & 0.900 \\
\bottomrule
\end{tabular}
\end{table}

\section{Discussion}
\subsection{Relationship to Prior Findings}
The results do not reconcile the findings of Lumer et al.~\cite{lumer} and PageIndex~\cite{pageindex}, which are not opposing measurements of the same thing: the 68\% figure is the share of pairwise comparisons in which an LLM judge preferred the vector-based system's answer, whereas the 98.7\% figure is a reported FinanceBench accuracy for a different system. The two studies differ in systems, data and metrics, and this small comparison, whose configurations are themselves confounded, cannot show whether query complexity accounts for their different outcomes. Answering that question would need a matched evaluation of both approaches on the same queries.

\subsection{FinanceBench in Context of Prior Work}
The FinanceBench results (Tree 0.938, Hybrid AHR 0.901, Vector 0.821 on a continuous 0--1 scale) are not directly comparable with the broader FinanceBench literature. Prior vector-based systems report binary pass/fail accuracy of 19\%~\cite{financebench}, 25.6\%~\cite{setty}, 32.6\%~\cite{jimeno}, and 49.66\%~\cite{finsage} when retrieving from full filings, a progression driven largely by improved chunking and reranking. In our setting retrieval is restricted to the evidence snippets supplied with each question, which makes the task considerably easier and means the results test which supplied snippet is selected rather than navigation of complete filings. Testing the PageIndex hypothesis~\cite{pageindex} that reasoning-based retrieval benefits from the structure of full professional documents requires an evaluation on complete filings.

\subsection{The Case for Adaptive Retrieval}
The Hybrid AHR used vector search for 45\% of queries (10/22, avg.\ score 0.830) and tree reasoning for 55\% (12/22, avg.\ score 0.858). It did not assign any query to Tier~4, so the fusion path was not exercised, and its predicted tier counts (10, 8, 4, 0) differ from the annotated counts (8, 7, 6, 1), so at least three queries were assigned a different tier from their annotation. AHR's scores largely follow the retriever it selects: it matches Vector RAG's weak medical Tier~1 result (0.50) and shares Tree Reasoning's strength on cross-reference queries. Because AHR generated its answers in model calls separate from those of the fixed configurations, its score on a routed query need not equal the score of the corresponding fixed system. Whether a router that agrees more often with the tier annotations, or one that defaults to tree reasoning when uncertain, would improve answer quality is a hypothesis for future work under matched conditions.

\subsection{Implications for Enterprise Deployment}
For regulated industries (finance, healthcare, legal), the results are consistent with structure-aware retrieval being worth evaluating for precision-critical queries, where missing referenced evidence can have material consequences, such as understating drug safety signals~\cite{goel} or missing contractual obligations. The present evidence is not sufficient to recommend a default: it comes from answer ratings on a small corpus, with confounded configurations and no valid retrieval metrics.

\subsection{Limitations}
These results are preliminary. The controlled corpus contains four short documents written for this study, with 22 queries; Tier~4 contains a single query, and each domain other than finance is represented by one document. FinanceBench was evaluated on the first 50 public questions (10 of 32 companies) with retrieval restricted to the supplied evidence snippets. All results come from a single run, reported as means without confidence intervals or significance tests. GPT-4o-mini both generates and judges the answers, which may bias scores toward its own style, and the two evaluations use different judge prompts and scales. Tree Reasoning is given the query's target document while Vector RAG searches all documents, which favours Tree Reasoning, and in the controlled experiment the two also use different generation prompts and output limits (a concise answer capped at 300 tokens for Vector RAG and a detailed one capped at 400 for Tree Reasoning), so quality differences may reflect these settings rather than the retriever. Retrieval precision, recall and F1 are not reported because the original scorer ignored document identity (Section~3.4). The vector baseline uses neither reranking nor hybrid lexical search, and the router's tier accuracy was not measured directly.

\section{Future Work}
Five directions emerge: (1) Scaling to the full 150-question FinanceBench set over complete filings and 1,200+ real SEC filings, with real legal and medical corpora, matched prompts and output limits, equal document access, document-qualified retrieval scoring, repeated runs, and statistical significance testing. (2) Incorporating cross-encoder reranking and hybrid lexical search into the vector baseline, with document selection applied equally to all methods; Lumer et al.~\cite{lumer} report MRR@5 rising from 0.160 to 0.750 with reranking. (3) Developing reinforcement learning-based meta-retrieval agents that learn optimal routing from query-outcome feedback. (4) Extending to multimodal documents where charts and images require vision-language model integration~\cite{clip,layoutparser}. (5) Investigating privacy-preserving tree indexing for regulated industries~\cite{chatterjee}.

\section{Conclusion}
This paper presents a small-scale cross-domain experimental comparison of Vector RAG, Tree Reasoning, and Adaptive Hybrid Retrieval on financial, legal, and medical documents, with a supplementary evaluation on FinanceBench evidence snippets. Tree Reasoning receives the highest mean judge quality on the controlled corpus (0.900 against 0.845) and on FinanceBench (0.938 against 0.821 for Vector RAG), while Vector RAG is about twice as fast. No configuration leads on every tier: Vector RAG leads on financial documents and the single multi-document query, Tree Reasoning on legal and medical documents, and the Adaptive Hybrid router, which chooses a retriever from the question alone, does not outperform fixed tree reasoning overall. Because document access, prompts and output limits differ between configurations, these single-run observations motivate, rather than establish, a controlled comparison with matched settings and document-qualified retrieval scoring.

\section*{Acknowledgement}
This work was supported by the Tuwaiq Research and Development Center, Tuwaiq Academy.


\begin{thebibliography}{99}
\bibitem{lumer} E. Lumer, M. Melich, O. Zino, et al., ``Rethinking retrieval: From traditional RAG to agentic and non-vector reasoning systems in the financial domain for LLMs,'' arXiv:2511.18177, Nov. 2025.
\bibitem{pageindex} M. Zhang, Y. Tang, and PageIndex Team, ``PageIndex: Next-generation vectorless, reasoning-based RAG,'' VectifyAI, Sep. 2025.
\bibitem{gao} Y. Gao, Y. Xiong, X. Xu, et al., ``Retrieval-augmented generation for large language models: A survey,'' arXiv:2312.10997, 2024.
\bibitem{liu} N. F. Liu, K. Lin, J. Hewitt, et al., ``Lost in the middle: How language models use long contexts,'' TACL, vol. 12, pp. 157--173, 2024.
\bibitem{lewis} P. Lewis, E. Perez, A. Piktus, et al., ``Retrieval-augmented generation for knowledge-intensive NLP tasks,'' in Proc. NeurIPS, 2020.
\bibitem{dpr} V. Karpukhin, B. Oguz, S. Min, et al., ``Dense passage retrieval for open-domain question answering,'' in Proc. EMNLP, 2020.
\bibitem{sarkar} P. Sarkar, ``Proxy-pointer RAG: Achieving vectorless accuracy at vector RAG scale and cost,'' Towards Data Science, Apr. 2026.
\bibitem{nogueira} R. Nogueira and K. Cho, ``Passage re-ranking with BERT,'' arXiv:1901.04085, 2019.
\bibitem{sun} W. Sun, L. Yan, X. Ma, et al., ``Is ChatGPT good at search? Investigating large language models as re-ranking agents,'' in Proc. EMNLP, 2023.
\bibitem{ranjan} R. Ranjan et al., ``A comprehensive survey of retrieval-augmented generation: Evolution, current landscape and future directions,'' arXiv:2410.12837, 2024.
\bibitem{ma} X. Ma, Y. Gong, P. He, et al., ``Query rewriting for retrieval-augmented large language models,'' arXiv:2305.14283, 2023.
\bibitem{adaptiverag} S. Jeong, J. Baek, S. Cho, S. J. Hwang, and J. C. Park, ``Adaptive-RAG: Learning to adapt retrieval-augmented large language models through question complexity,'' in Proc. NAACL, 2024.
\bibitem{yan} S.-Q. Yan, J.-C. Gu, Y. Zhu, and Z.-H. Ling, ``Corrective retrieval augmented generation,'' arXiv:2401.15884, 2024.
\bibitem{hu} Y. Hu and Y. Lu, ``RAG and RAU: A survey on retrieval-augmented language model in natural language processing,'' arXiv:2404.19543, 2024.
\bibitem{wu} S. Wu, Y. Xiong, Y. Cui, et al., ``Retrieval-augmented generation for natural language processing: A survey,'' arXiv:2407.13193, 2024.
\bibitem{zhao} P. Zhao et al., ``Retrieval-augmented generation for AI-generated content: A survey,'' arXiv:2402.19473, 2024.
\bibitem{sharma} C. Sharma, ``Retrieval-augmented generation: A comprehensive survey of architectures, enhancements, and robustness frontiers,'' arXiv:2506.00054, 2025.
\bibitem{yu} H. Yu, A. Gan, K. Zhang, et al., ``Evaluation of retrieval-augmented generation: A survey,'' arXiv:2405.07437, 2024.
\bibitem{colbert} O. Khattab and M. Zaharia, ``ColBERT: Efficient and effective passage search via contextualized late interaction over BERT,'' in Proc. SIGIR, 2020.
\bibitem{bm25} S. Robertson and H. Zaragoza, ``The probabilistic relevance framework: BM25 and beyond,'' Found. Trends Inf. Retr., vol. 3, no. 4, pp. 333--389, 2009.
\bibitem{setty} S. Setty, H. Thakkar, A. Lee, et al., ``Improving retrieval for RAG based question answering models on financial documents,'' arXiv:2404.07221, 2024.
\bibitem{ragfusion} Z. Rackauckas, ``RAG-Fusion: A new take on retrieval-augmented generation,'' arXiv preprint, 2024.
\bibitem{alphago} D. Silver, A. Huang, C. J. Maddison, et al., ``Mastering the game of Go with deep neural networks and tree search,'' Nature, vol. 529, no. 7587, pp. 484--489, 2016.
\bibitem{graphrag} D. Edge, H. Trinh, N. Cheng, et al., ``From local to global: A graph RAG approach to query-focused summarization,'' arXiv:2404.16130, 2024.
\bibitem{hirag} H. Huang, Y. Huang, J. Yang, et al., ``Retrieval-augmented generation with hierarchical knowledge,'' in Findings of EMNLP, 2025.
\bibitem{got} M. Besta, N. Blach, A. Kubicek, et al., ``Graph of thoughts: Solving elaborate problems with large language models,'' in Proc. AAAI, 2024.
\bibitem{kragen} N. Matsumoto et al., ``KRAGEN: A knowledge graph-enhanced RAG framework for biomedical problem solving using large language models,'' 2024.
\bibitem{tatqa} F. Zhu, W. Lei, Y. Huang, et al., ``TAT-QA: A question answering benchmark on a hybrid of tabular and textual content in finance,'' in Proc. ACL, 2021.
\bibitem{finsage} X. Wang, J. Chi, Z. Tai, et al., ``FinSage: A multi-aspect RAG system for financial filings question answering,'' arXiv:2504.14493, 2025.
\bibitem{rlfkv} ``Mitigating hallucination in financial retrieval-augmented generation via fine-grained knowledge verification,'' arXiv:2602.05723, 2026.
\bibitem{rankrag} Y. Yu et al., ``RankRAG: Unifying context ranking with retrieval-augmented generation in LLMs,'' in Proc. NeurIPS, 2024.
\bibitem{sbert} N. Reimers and I. Gurevych, ``Sentence-BERT: Sentence embeddings using Siamese BERT-networks,'' in Proc. EMNLP, 2019.
\bibitem{faiss} J. Johnson, M. Douze, and H. J\'egou, ``Billion-scale similarity search with GPUs,'' IEEE Trans. Big Data, vol. 7, no. 3, pp. 535--547, 2021.
\bibitem{zheng} L. Zheng, W.-L. Chiang, Y. Sheng, et al., ``Judging LLM-as-a-judge with MT-Bench and Chatbot Arena,'' in Proc. NeurIPS, 2023.
\bibitem{goel} G. Goel, ``PageIndex: The vectorless RAG,'' PVTech Substack, Mar. 2026.
\bibitem{clip} A. Radford, J. W. Kim, et al., ``Learning transferable visual models from natural language supervision,'' in Proc. ICML, 2021.
\bibitem{layoutparser} Z. Shen, R. Zhang, M. Dell, et al., ``LayoutParser: A unified toolkit for deep learning based document image analysis,'' in Proc. ICDAR, 2021.
\bibitem{chatterjee} S. Chatterjee, ``The structural pivot: Analytical perspectives on vectorless RAG and hierarchical page indexing,'' Medium, Feb. 2026.
\bibitem{financebench} P. Islam, A. Kannappan, D. Kiela, R. Qian, N. Scherrer, and B. Vidgen, ``FinanceBench: A new benchmark for financial question answering,'' arXiv:2311.11944, 2023.
\bibitem{jimeno} A. Jimeno-Yepes, Y. You, J. Milczek, S. Laverde, and R. Li, ``Financial report chunking for effective retrieval augmented generation,'' arXiv:2402.05131, 2024.
\end{thebibliography}
\end{document}